\documentclass[10pt,a4paper]{article}
\usepackage[nofillcomment,vlined,linesnumbered,titlenumbered,algo2e,ruled]{algorithm2e}
\usepackage{listings}
\usepackage[]{graphicx}
\usepackage{makecell}

\newcommand{\sImagePath}{./}

\title{A Fast and Generic GPU-Based Parallel Reduction Implementation}

\begin{document}

\maketitle

\author{Walid Jradi$^{\rm a}$$^{\ast}$\footnote{$^\ast$Corresponding author. e-mail: walid.jradi@gmail.com \vspace{6pt}}, Hugo do Nascimento and Wellington Martins\\\vspace{6pt}  $^{\rm a}${\em{Universidade Federal de Goi\'{a}s -- Instituto de Inform\'{a}tica\\Alameda Palmeiras, Quadra D, Campus Samambaia\\CEP 74690-900 -- Goi\^{a}nia -- Goi\'{a}s}} }

\begin{abstract}
Reduction operations are extensively employed in many computational problems. A reduction consists of, given a finite set of numeric elements, combining into a single value all elements in that set, using for this a combiner function. A parallel reduction, in turn, is the reduction operation concurrently performed when multiple execution units are available. The current work reports an investigation on this subject and depicts a GPU-based parallel approach for it. Employing techniques like \textit{Loop Unrolling}, \textit{Persistent Threads} and \textit{Algebraic Expressions} to avoid thread divergence, the presented approach was able to achieve a 2.8x speedup when compared to ???, using a generic, simple and easily portable code. Experiments conducted to evaluate the approach show that the strategy is able to perform efficiently in AMD and NVidia's hardware, as well as in OpenCL and CUDA.

\end{abstract}

\section{Introduction}
\label{sec:Introduction}

Widely used as a basic subroutine for a number of algorithms such as Counting Sort~\cite{Cormen2002}, Stream Compaction~\cite{Billeter:2009:ESC:1572769.1572795}, Golden Section and Fibonacci Methods~\cite{Kiefer1953} and Radix Sort~\cite[Chapter 8.3]{Cormen2002}.

The remainder of this paper is structured as follows. Section~\ref{subsec:PR_ProblemDefinition} presents some basic concepts. Section~\ref{sec:GPUParallelReduction} briefly describes the techniques currently in use. Section~\ref{sec:NewApproachPR} explains our approach. Section~\ref{sec:computationalExperimentsPR} details the experimental environment and the results. Finally, Section~\ref{sec:GeneralRemarksParallelReduction} gives general remarks about the presented strategy.

\subsection{Problem Definition}
\label{subsec:PR_ProblemDefinition}

Formally, a reduction can be defined as follows~\cite{Parhami1999}: Given a set \textit{X} with \textit{n} values, $X = \lbrace x_0, x_1, ..., x_{n-1} \rbrace$, compute $x_0 \otimes x_1 \otimes ... \otimes x_{n-1}$. The associative operator $\otimes$ (also known as \textit{combiner function}) can be (but is not limited to) any one of the set $\lbrace +, \times, \wedge, \vee, \oplus, \cap, \cup, \max, \min \rbrace$.

Consider the pseudo code shown in Algorithm~\ref{alg:Summation}. At first glance, it seems that the algorithm is inherently sequential, since the variable \textit{accumulator} depends on the value computed in the previous step, preventing any attempt of parallelization. However, it is possible to avoid this problem by making use of two basic properties of addition and multiplication operations: \textit{Associativity} and \textit{Commutativity}\footnote{Other two properties, \textit {Neutral Element} and \textit {Closeness}, guarantee, respectively, that any number added to zero results in the number itself, and when we add/multiply two or more numbers within the same set (natural, for example), the result will always be a number within the same set.}.

\begin{algorithm2e}[ht]
 \DontPrintSemicolon % Some LaTeX compilers require you to use \dontprintsemicolon instead
 \KwIn{A set $A=\{a_1, a_2, \ldots, a_n\}$ of numeric elements}
 \KwOut{The sum of all elements}
 
 $accumulator \gets 0$\;
 \For{$i \gets 1$ \textbf{to} $n$}
 {
  $accumulator \gets accumulator + a_i$\;
 }
 \Return{$accumulator$}\;
 \caption{\textit{Summation(A)}}
 \label{alg:Summation}
\end{algorithm2e}

\begin{itemize}

 \item \textbf{Associativity} means that, given three or more numbers, they can be linked in any order without changing the final result. Taking the sum as an example, it's possible to do $a_1 + a_2$ and, then, add $a_3$, and the result will be the same as doing $a_3 + a_2$ and then adding $a_1$. Formally, we have $(a_1 + a_2) + a_3 \equiv a_1 + (a_2 + a_3)$;
 
 \item \textbf{Commutativity} ensures that no matter the order in which an operation on two numbers $a_1$ and $a_2$ is performed, the result will always be the same. Formally, for multiplication, we have $a_1 \cdot a_2 \equiv a_2 \cdot a_1$.

\end{itemize}

\begin{figure}[ht]
  \centering
  \includegraphics[width=1.0\textwidth]{\sImagePath 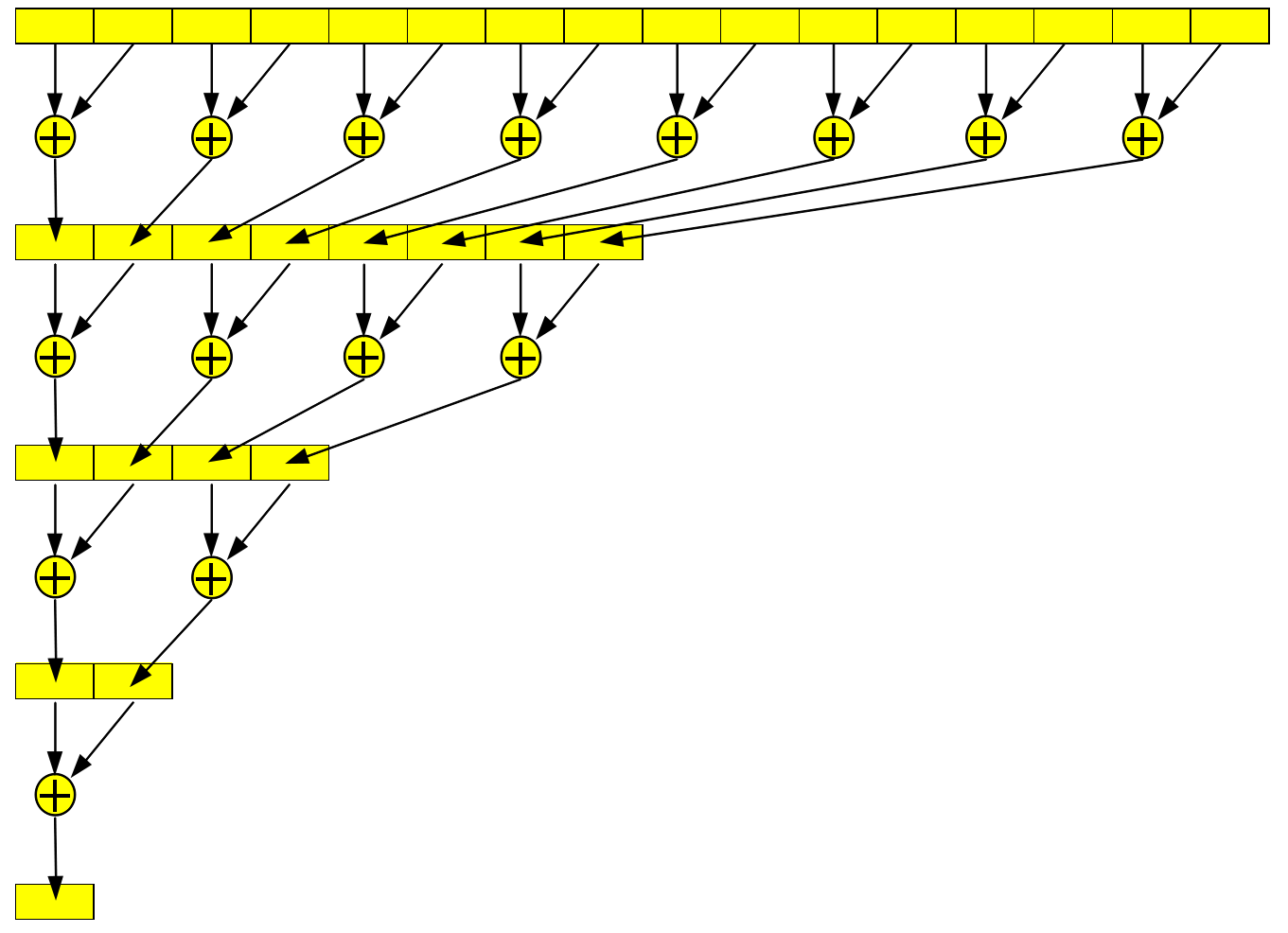}
  \caption{Parallel reduction -- associative reduction tree.}
  \label{fig:AssociativeReductionTree}
 \end{figure}

Considering that the order in which the elements are combined does not affect the final result\footnote{Although, mathematically, this is true for numbers in any set, in computational terms things are a little more complicated. For instance, these properties hold for the set of integers, but the same does not happen for the floating point numbers due to the inherent imprecision that arises when combining (adding, multiplying, etc.) numbers with different exponents, which leads to the absorption of the lower bits during the combine operation. As an example, mathematically the result of $(1.5 + 4^{50} - 4^{50})$ is always the same, no matter the order the terms are added, whereas the floating point computed value can result in 0 or 1.5, depending on the sequence in which operations are performed~\cite{Defour7321514, Goldberg:1991:CSK:103162.103163, higham2002accuracy, muller2009handbook}.}\textsuperscript{, }\footnote{Note that, although this is a complicating factor when a large numerical precision is necessary, it did not actually preclude its application in a problem when the accumulated error using single-precision floats did not exceed a certain pre-defined threshold. On the other hand, if such precision becomes necessary, the problem could be greatly minimized by adopting the use of double-precision floating points (which potentially can decrease the application performance for certain GPU models) or using some strategies to reduce truncation errors, like the one proposed by Kahan~\cite{Kahan:1965:PRR:363707.363723}, among others.}, these two properties can be used, dividing the problem into smaller subproblems and these, in turn, solved in parallel. After solving each subproblem, the partial results are combined to produce the final result. Figure~\ref{fig:AssociativeReductionTree} illustrates the process using the associative operator ``$+$'' in an array with 16 elements.

\section{Parallel Reduction in GPUs}
\label{sec:GPUParallelReduction}

Since the arrival of programmable GPUs, some strategies to accelerate the reduction operation on such devices have been proposed. The two most well known are those described by Mark Harris~\cite{harris2007optimizing} and Bryan Catanzaro~\cite{AMDParallelReduction2014}. Most recently, Justin Luitjens~\cite{JustinLuitjens2014} presented some improvements to the strategies described in~\cite{harris2007optimizing}. Unfortunately, the strategies adopted by~\cite{harris2007optimizing} and~\cite{JustinLuitjens2014}, although very efficient, are limited to hardware and software provided by NVidia, restricting their use.

On the other hand, the proposal of Catanzaro~\cite{AMDParallelReduction2014} is based on the open standard OpenCL~\cite{Khronos2008OpenCL}, adopted by a myriad of manufacturers, what makes it portable. Nevertheless, the code presented in~\cite{AMDParallelReduction2014} also has a weakeness, as it does not adopt some strategies that could significantly improve its performance.

This section details how the \textit{associative} and \textit {commutative} properties can be used to implement efficient parallel reductions on GPUs. As highlighted at the end of Section~\ref{subsec:PR_ProblemDefinition}, the basic idea is to ``split'' the problem into smaller pieces and solve them in parallel. However, the execution environment (GPU hardware) imposes some restrictions that must be considered to maximize the \textit{speedup}. Therefore, the details of how GPUs are organized~\cite{Che20081370, wilt2013cuda} will dictate the choices from now on.

The approaches of Harris~\cite{harris2007optimizing} and Catanzaro~\cite{AMDParallelReduction2014} to deal with reductions in GPUs operate in a pretty similar way, using a tree-based structure.

One of the aspects to be considered is the number of elements in the collection (vector) in which the reduction will be applied. If this amount is sufficiently small and can be stored in the local memory of each SM, then the reduction becomes quite simple. In~\cite{AMDParallelReduction2014}, Catanzaro presents some strategies for this case and conducts performance comparisons between them. Then, after describing how reductions can be efficiently performed in small sets, Catanzaro shifts his focus to the cases in which a large volume of data must be handled. Three strategies are presented and a winner, called ``\textit{Two-Stage Parallel Reduction}'', is elected. Harris~\cite{harris2007optimizing} deals only with parallel reduction in large datasets.

Our approach is mainly based on a proposal from Catanzaro~\cite{AMDParallelReduction2014}. Therefore, a more detailed description of it is presented. First, however, we also give an explanation of the strategies by Harris~\cite{harris2007optimizing} and Luitjens~\cite{JustinLuitjens2014}, since some ideas for speeding up the computation came from them. Hence, unlike the rest of the thesis, here their original code is presented, and not just the pseudo code.

\subsection{Mark Harris' Work}
\label{subsec:MarkHarris}

The work presented by Harris~\cite{harris2007optimizing} focuses on techniques for performing reductions of large data volumes. The author shows, through successive versions of the same algorithm, how bad decisions or an incorrect way of mapping the problem to the target platform can negatively impact the application performance.

Problems like shared memory bank conflict, lack of communication between thread blocks (making it impossible for a kernel to reduce a large array at once) and highly divergent warps are addressed. Starting with a naive version, step by step improvements are described, reaching an implementation 30x faster than the first one. Next, we show how the author achieved such speedups.

Harris performed experiments using a G80 GPU. This video card has a 384-bit memory interface, with a 900 MHz DDR memory, which leads to a theoretic $\frac{384 \ast 1800}{8} = 86.4 GB/s$ of memory bandwidth\footnote{Memory bandwidth basically determines how fast is the memory. Usually, it is measured in gigabytes per second (GB/s). The more bandwidth of the memory and the more it is explored by the running program, the faster the computation.}. All tests were conducted using a vector with $2^{22}$ (4M) integer values.

As a result of all the applied optimizations, the final version of the code runs in 0.268ms and the memory bandwidth usage reaches 62.671GB/s. All these improvements are summarized in Table~\ref{table:ExecutionTimesMH}.

\begin{table*}[ht]
 \centering
  \begin{tabular}{|p{7cm}|c|c|c|c|}
  \hline
								& \thead{Time \\ (ms)}	& \thead{Memory \\ Bandwidth \\ (GB/s)}	& \thead{Step \\ speedup}& \thead{Cummulative \\ speedup}\\ \hline \hline
Kernel 1: interleaved addressing with divergent branching	&	8.054	&	2.083			 	&			&				\\ \hline
Kernel 2: interleaved addressing with bank conflicts		&	3.456	&	4.854				&	2.33x		&	2.33x			\\ \hline
Kernel 3: sequential addressing					&	1.722	&	9.741				&	2.01x		&	4.68x			\\ \hline
Kernel 4: first add during global load				&	0.965	&	17.377				&	1.78x		&	8.34x			\\ \hline
Kernel 5: unroll last warp					&	0.536	&	31.289				&	1.8x		&	15.01x			\\ \hline
Kernel 6: completely unrolled					&	0.381	&	43.996				&	1.41x		&	21.16x			\\ \hline
Kernel 7: multiple elements per thread				&	0.268	&	62.671				&	1.42x		&	30.04x			\\ \hline
  \end{tabular}
 \caption{Performance for parallel reduction of $2^{22}$ integer elements (extracted from~\cite{harris2007optimizing}).}
 \label{table:ExecutionTimesMH}
\end{table*}

\subsection{Justin Luitjens' Work}
\label{subsec:JustinLuitjensPR}

In~\cite{JustinLuitjens2014} Luitjens shows how a new feature of the NVidia's Kepler (and newer) GPU architecture can be used to make reductions even faster when compared to the strategies presented in~\cite{harris2007optimizing}: the shuffle (SHFL) instruction.

Usually, work-items inside the same SM use the local (shared) memory when they need to communicate (exchange information). This involves a three-step process: writing the data to local memory, perform a synchronization barrier and then read the data back from local memory. The Kepler and newer architectures implement the \textit{shuffle} instruction, which enables a work-item to directly read private data from another work-item in the same wave-front. According to the author, there are four main advantages in using this instruction:

\begin{itemize}

 \item It ultimately allows work-items inside a wave-front to collectively exchange or broadcast data;

 \item It replaces the three-step process by a single instruction, effectively increasing the bandwidth and decreasing the latency;
 
 \item It does not use the local memory at all;
 
 \item A sync barrier is implicit in the instruction and, hence, a synchronization step inside a workgroup is not necessary.

\end{itemize}

\begin{figure}[ht]
  \centering
  \includegraphics[width=0.6\textwidth]{\sImagePath 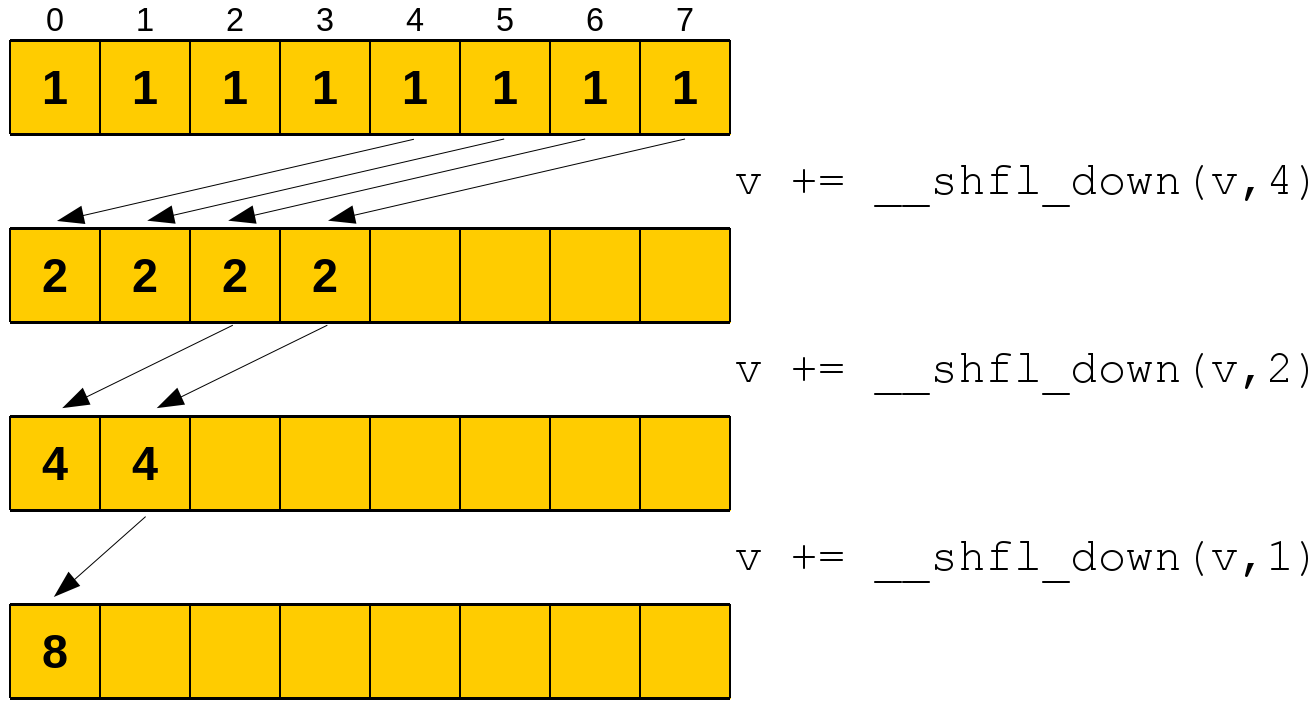}
  \caption{Parallel reduction using the shuffle instruction (extracted from~\cite{JustinLuitjens2014}).}
  \label{fig:ParallelReduction_Shuffle_Instruction}
 \end{figure}

Figure~\ref{fig:ParallelReduction_Shuffle_Instruction} shows how this instruction can be used to build a reduction tree. As pointed out by Luitjens, this figure only includes the arrows for the work-items actually doing useful work. All work-items are indeed shifting values even though these values are not necessary in the reduction process.

Using this instruction, several versions of the reduction were proposed, implemented and compared. However, although Luitjens states that the adopted strategies lead to faster reductions than those described by Harris~\cite{harris2007optimizing}, no comparative studies between the two approaches were conducted. 

\subsection{Bryan Catanzaro's Work}
\label{subsec:BryanCatanzaroPR}

% In~\cite{AMDParallelReduction2014}, after describing how reductions can be efficiently performed in small sets, the author shifted his focus to the cases in which a large volume of data must be handled. Three strategies are presented and a winner, called ``\textit{Two-Stage Parallel Reduction}'', is elected. Next, we briefly describe how it works.

% and how, with minor conceptual changes, its running time can be significantly reduced.

Now, we describe Catanzaro's two-stage parallel reduction approach for large datasets, as presented in~\cite{AMDParallelReduction2014}.

The technique is based on dividing the data set in \textit{p} pieces (or ``\textit{chunks}''), where \textit{p} is large enough to keep all GPU cores busy. It is also necessary to limit the number of \textit{work-items} to the maximum amount that the GPU can handle in total without having to switch between them (from now on, that maximum will be called \textit{GS} -- or \textit{global size}). Each chunk is then processed by a work-group.

Since the sum operation has the properties of associativity and commutativity, each \textit{work-item} can perform its own reduction sequentially and intercalary with the others. A work-item takes, as the starting point, its global identifier and accumulates, in a private variable, its partial sum, skipping \textit{GS} positions at every step in the vector stored in the GPU's global memory.

After having completed a pass through the data set, the \textit{work-items} in each workgroup write the result of their own reduction in a scrap vector located in local/shared memory which, in turn, will also be reduced in parallel. At the end of the process, each working group will have its own scrap containing, in its position 0, the result of the reduction so far. This partial result is then copied to another vector, this time stored in the GPU global memory, which size must be equal to |SM|. The first stage is then complete. Its source code, extracted from~\cite{AMDParallelReduction2014}, is presented in Listing~\ref{code:ParallelReductionBC1}.

\noindent
\begin{minipage}{\linewidth}
\begin{lstlisting}[escapechar=\|,basicstyle=\footnotesize,firstnumber=1,frame=single,language=C,caption=Two-stage parallel reduction of Catanzaro -- stage 1\label{code:ParallelReductionBC1}]
__kernel void reduce(__global float* buffer,
            __local float* scratch,
            __const int length,
            __global float* result) {

  int global_index = get_global_id(0); |\label{ln040}|
  float accumulator = INFINITY; |\label{ln041}|
  // Loop sequentially over chunks of input vector
  while (global_index < length) { |\label{ln042}|
    float element = buffer[global_index]; |\label{ln043}|
    accumulator = (accumulator < element) ? accumulator : element; |\label{ln044}|
    global_index += get_global_size(0); |\label{ln045}|
  }
  int local_index = get_local_id(0); |\label{ln046}|
  scratch[local_index] = accumulator; |\label{ln047}|
  barrier(CLK_LOCAL_MEM_FENCE); |\label{ln048}|
  // Perform parallel reduction
  for(int offset=get_local_size(0)/2; offset>0; offset=offset/2) { |\label{ln049}|
    if (local_index < offset) { |\label{ln050}|
      float other = scratch[local_index + offset]; |\label{ln051}|
      float mine = scratch[local_index]; |\label{ln052}|
      scratch[local_index] = (mine < other) ? mine : other; |\label{ln053}|
    }
    barrier(CLK_LOCAL_MEM_FENCE); |\label{ln054}|
  }
  if (local_index == 0) { |\label{ln055}|
    result[get_group_id(0)] = scratch[0]; |\label{ln056}|
  }
}
\end{lstlisting}
\end{minipage}

The second stage is simpler. Since now there is a vector with $|SM|$ elements in the global memory -- with the result of a partial sum in each position -- just the first $|SM|$ \textit{work-items} of the first $SM$ copy their corresponding value to an array allocated in local memory. Then the \textit{work-items} perform a new parallel sum of the elements in the vector. After copying the value in position 0 back to global memory, the reduction is finally complete. 

%Figures~\ref{fig:ParallelReduction_Stage1_Step_1} to~\ref{fig:ParallelReduction_Stage2_Single_Step} illustrate these two stages, assuming the presence of two SMs on the GPU, each one able to run four \textit{work-items}.
%
% \begin{figure}
%  \centering
%  \includegraphics[width=0.775\textwidth]{fig/ParallelReduction_Stage1_Step1.png}
%  \caption{Parallel reduction -- first stage, step 1.}
%  \label{fig:ParallelReduction_Stage1_Step_1}
% \end{figure}
%
% \begin{figure}
%  \centering
%  \includegraphics[width=0.3\textwidth]{fig/ParallelReduction_Stage1_Step2.png}
%  \caption{Parallel reduction -- first stage, step 2.}
%  \label{fig:ParallelReduction_Stage1_Step_2}
% \end{figure}
%
% \begin{figure}
%  \centering
%  \includegraphics[width=0.2\textwidth]{fig/ParallelReduction_Stage1_Step3.png}
%%   \caption[Parallel reduction: first stage, step 3.]{Parallelreduction: first stage, step 3. Each work-item performs its own local reduction.}
%  \caption[Parallel reduction -- first stage, step 3.]{Parallel reduction -- first stage, step 3.}
%  \label{fig:ParallelReduction_Stage1_Step_3}
% \end{figure}
%
% \begin{figure}
%  \centering
%  \includegraphics[width=0.275\textwidth]{fig/ParallelReduction_Stage1_Step4.png}
%  \caption{Parallel reduction -- first stage, step 4.}
%  \label{fig:ParallelReduction_Stage1_Step_4}
% \end{figure}
%
% \begin{figure}
%  \centering
%  \includegraphics[width=0.8\textwidth]{fig/ParallelReduction_Stage2_Step1.png}
%  \caption{Parallel reduction -- second stage, single step.}
%  \label{fig:ParallelReduction_Stage2_Single_Step}
% \end{figure}

The next sections detail some advanced techniques to further explore parallelism and that were extensively used in the present work.

\subsection{Loop Unrolling}
\label{subsec:LoopUnrolling}

%https://software.intel.com/sites/landingpage/opencl/optimization-guide/Note_on_Loops.htm
%http://stackoverflow.com/questions/7703443/inter-block-barrier-on-cuda

\textit{Loop Unrolling} (also known as \textit{Loop Unwinding} and \textit{Loop Unfolding}) is an optimization technique -- performed by the compiler or manually by the programmer -- applicable to certain kinds of loops in order to reduce (or even prevent) the occurrence of execution branches and minimize the cost of instructions for controlling the \textit{loop}~\cite{abrash1997michael, fog2008optimizing, Huang97generalizedloopunrolling, Sarkar:2001:OUN:564965.564968}. Its goal is to optimize the program's execution speed at the expense of increasing the size of the generated code (\textit{space-time tradeoff}). It is easily applicable to loops where the number of executions is previously known, like routines of vector manipulation where the number of elements is fixed.

Basically the technique consists in the reuse of the sequence of instructions being executed within the loop, so as to include more of an iteration of the code every time the \textit{loop} is repeated, reducing the amount of these repetitions.

This reuse is done by manually replicating the code inside the \textit{loop} a certain amount of times or through the ``\textit{\#pragma unroll \textbf{n}}''\footnote{A \textit{directive pragma} is a language construct that provides additional information to the compiler, specifying how to process its input. This additional information usually is beyond what is conveyed in the language itself.} positioned immediately before the beginning of the loop. The number of times the loop is unrolled is called \textit{Unrolling Factor} and, with the pragma directive, it is given by the parameter ``\textit{\textbf{n}}''.

It is worth noting that with the pragma directive we leave the decisions of how the loop should be unrolled to the compiler, which may lead to a not so optimized resulting code. In the experiments performed as part of this work, the best results were always achieved using manual loop unrolling.

As an example, consider the C code shown in Listing~\ref{code:VectorMultiply1}, which simply multiplies the elements of an array by its index ($a_i \leftarrow a_i \cdot i $). In this example, we call \textit{\textbf{L}} the \textit{loop size} and \textit{\textbf{F}} its \textit{unrolling factor}. \textit{\textbf{L}} here is equal to 100.

\begin{lstlisting}[frame=single,floatplacement=H,language=C,caption=Multiplying elements in a vector\label{code:VectorMultiply1}]
for (int i = 0; i < 100; i++) {
a[i] = a[i] * i;
}
\end{lstlisting}

It's possible to significantly improve the execution speed of this algorithm by unrolling it, as shown in Listing~\ref{code:VectorMultiply2}.

\begin{lstlisting}[frame=single,floatplacement=H,language=C,caption=Unrolling the multiply routine\label{code:VectorMultiply2}]
for (int i = 0; i < 100; i += 3) {
a[i]   = a[i]*i;
a[i+1] = a[i+1]*(i+1);
a[i+2] = a[i+2]*(i+2);
}
\end{lstlisting}

The two extra lines of code and the ``\textit{i += 3}'' in Listing~\ref{code:VectorMultiply2} performs the desired three-fold ($F = 3$) manual loop unrolling.

As it can be seen, the $\frac{L}{F}$ ratio does not necessarily need to be an integer. If it admits a remainder, the compiler can (since the number of iterations is previously known at compile time) add extra code to the end of the unrolled generated code in order to ensure its correctness.

Unrolling, when applicable, offers several advantages over non-unrolled code. Besides the decrease in the number of iterations, an increase occurs in the amount of work done each time through the loop. This also open ways for the exploration of parallelism by the compiler in machines with multiple execution units, since each instruction within the \textit{loop} can be handled by an independent thread.

However, these are only the most easily perceivable benefits. Agner Fog~\cite{fog2008optimizing} listed several others, as well as some observations about when this technique should be avoided. Such factors (advantages and disadvantages) must be considered by the programmer when deciding to use loop unrolling or not.

\subsection{Persistent Threads}
\label{subsec:PersistentThreads}

Since the launch of the first programmable GPUs and with all its basic architecture inspired by the SIMD model, the ``\textit{Single Instruction Multiple Thread''} (SIMT) and ``\textit{Single Program Multiple Data}'' (SPMD)  paradigms have become standards \textit {de facto}. Both seek to hide the details of the underlying \textit{hardware} where the code runs, attempting to facilitate the painful task of development~\cite{gupta2012study}.

Gupta et al.~\cite{gupta2012study} argue that the usage of these ``traditional'' paradigms greatly limits the actions of the programmer, because all control of the execution flow is in the power of the \textit{scheduler's} video card. This programming style, which delegates all the decisions to the scheduler, is called by the authors as ``non-PT'', or ``non-Persistent''.

It requires that the software developer abstracts units of work to virtual work-items. Since the number of wave-fronts to create is based on the number of virtual work-items, during a kernel launch usually there are several hundreds of even thousands more wave-fronts to be executed than the amount of physical processing elements to assign them to.

Such scheduling of wave-fronts is performed by the \textit{scheduler} and the programmer has no means to interfere in the process, e.g., \textit{how}, \textit{where}, \textit{when} and in \textit{which order} the work-groups will be assigned.

Gupta et al. claim that, while these abstractions reduce the effort for new developers in the GPGPU field, they also create obstacles for experienced programmers, who normally face problems for which workload is inherently irregular, therefore making it much more difficult to efficiently parallelize when compared to problems whose parallel solution is more regular.

According to Gupta et al., this uncovers a serious drawback of the current SPMD programming style, which is not able to ensure \textit{order}, \textit{location} and \textit{timing}. It also does not allow the software developer to regulate these three parameters without completely avoiding the GPU scheduler.

Thus, to overcome these limitations, developers have been using a programming style called \textit{Persistent Threads (``PT'')}, whose low level of abstraction allows performance gains by directly controlling the scheduling of work-groups. And although this style has been in use for some time, only in 2012 it was formally introduced, described and analyzed by Gupta et al.~\cite{gupta2012study}. They also list several problems when adopting the traditional style.

Basically, what the PT style change is the \textit{lifetime} of a \textit{work-item}~\cite{6569834}, by letting it keep running longer and giving it much more work than in the traditional ``non-PT'' style~\cite{Steinberger:2012:SDS:2366145.2366180}. This is done circumscribing the logic kernel (or part of it) in a loop, so this loop remains running while there are items to be processed.

Briefly, from the point of view of the developer, all work-items are active while the kernel is running. As a direct consequence of PT, a \textit{kernel} should be triggered using only the amount of \textit{work-items} that can be executed concurrently by each Streaming Multiprocessor. All these actions will prevent constant return of control to the host and the cost of new kernel invocations~\cite{6569834}.

Gupta et al. acknowledge, however, that the technique of Persistent Threads is not a panacea, and its use should be carefully evaluated~\cite{gupta2012study}. In particular, the technique fits well when the amount of memory accesses is limited (i.e., few reading/writing to global memory and a large volume of computation) and the problem being solved has not many initial input elements or the growth in the number of elements in the input set is fairly limited. Beyond these conditions, the traditional non-PT style tends to outperform the PT style.

\subsection{Thread Divergence}
\label{subsec:AvoidingThreadDivergence}

Current GPUs are able to deliver massive computational power at a reasonably low cost. However, due to the way they are constructed, some obstacles must be overcome for the effective use of such power. One of the main and hardest obstacles to avoid is the presence of conditional statements~\cite{Zhang:2010:SGA:1810085.1810104} potentially leading to branches in the execution flow of the various work-items~\cite{Han:2011:RBD:1964179.1964184}.

By default, GPUs try to run all the work-items inside the wave-fronts in the SIMD model. However, if the code being executed has conditional statements that lead to divergences in program flow, the divergent work-items will be stalled and its execution will only happen after the non-stalled work-items have completed their runs, which ultimately compromises the desired \textit{speedup}. This phenomenon is called \textit{Thread Divergence}~\cite{CPE:CPE2931, Han:2011:RBD:1964179.1964184, Narasiman:2011:IGP:2155620.2155656, Zhang:2010:SGA:1810085.1810104}.

Trying to circumvent this problem, some strategies have been proposed in order to minimize or even eliminate the effects of such phenomena. Among them, we cite~\cite{CPE:CPE2931, 4408272, Han:2011:RBD:1964179.1964184, Meng:2010:DWS:1816038.1815992, Narasiman:2011:IGP:2155620.2155656, Zhang:2010:SGA:1810085.1810104}.

% This phenomena emerged during the implementation of the urban traffic assignment algorithm described in Chapter~\ref{chapter:traffic_assignment_problem}, in the step of determining the arc cost functions. Wherefore it became necessary to develop a method to prevent the flow divergence, which could ultimately compromise the performance of such a step of computation. The method is detailed at the end of Section~\ref{sec:NewApproach}.

---:::

Wherefore it became necessary to develop a method to prevent flow divergence, which could ultimately compromise the performance of such a step of computation. The method is detailed at the end of Section~\ref{sec:NewApproachPR}.

\section{The New Approach}
\label{sec:NewApproachPR}

The improvements proposed in our work focus on Steps 1 and 3 of the first stage of the reduction presented in Section~\ref{subsec:BryanCatanzaroPR}. The improvements employ the same strategies proposed by Harris~\cite{harris2007optimizing} to increase the performance of the approach originally presented by Catanzaro~\cite{AMDParallelReduction2014} but with appropriately chosen interventions.

In step 1 of the original implementation, the vector in global memory containing the data to be reduced is entirely traversed by the \textit{work-items}, each one performing its own reduction.

This step already uses the ``Persistent-Thread'' strategy, but its performance can be improved by adopting loop unrolling (Section~\ref{subsec:LoopUnrolling}). As it can be seen, instead of doing the unroll when the data is in local memory, as proposed by Harris~\cite{harris2007optimizing} (Listings~\ref{code:ParallelReductionMH5} and~\ref{code:ParallelReductionMH6} of Section~\ref{subsec:MarkHarris}), our improvement performs the unroll in the global memory.

The code presented in Listing~\ref{code:UnrollingStep1} shows the modified loop, assuming an unrolling factor (F) equals to 4, $iGlobalID$ as the \textit{work-item} global identifier and $iLength$ as the number of elements to be reduced.

\noindent
\begin{minipage}{\linewidth}
\begin{lstlisting}[frame=single,language=C,caption=Unrolling the step 1\label{code:UnrollingStep1}]
 for (iPos = iGlobalID*iUnrollingFactor; iPos < iLength;
      iPos += iGlobalSize*iUnrollingFactor)
 {
  i0 = iPos;   i1 = iPos+1; i2 = iPos+2; i3 = iPos+3;
  accumulator +=
  ((i0<iLength)*(aVector[i0])+
   (i1<iLength)*(aVector[i1])+
   (i2<iLength)*(aVector[i2])+
   (i3<iLength)*(aVector[i3]));
 }
\end{lstlisting}
\end{minipage}

A special attention must be given to how the data is brought from the global memory (\textit{aVector}) to the private memory (\textit{accumulator}), through the use of algebraic expressions that prevent reading from invalid memory locations, thus avoiding the usage of ``ifs'' and potential divergences in the execution flow. The expression $i_n < iLength $ expands to integers 1 or 0 whether it is, respectively, true or false. In the first case ($i_n < iLength) * (aVector [i_n])$ is interpreted as $(1) * (aVector [i_n])$, adding the value stored in location $i_n$ to the partial sum (\textit{accumulator}). In the second case, the expression is interpreted as $(0) * (aVector[0])$, ensuring that -- regardless of the data stored in the first position of the vector -- value 0 is added to \textit{accumulator}, keeping the partial sum correctness.

At the begining of Step 3, the resulting values of the previous sums are already stored in the local memory of the SMs. Then, each SM performs its own local reduction with its work-items.

In the solutions presented by Harris~\cite{harris2007optimizing} and Catanzaro~\cite{AMDParallelReduction2014}, in this step all \textit{work-items} are kept synchronized through the use of barriers. However, with minor conceptual changes, it is possible to completely eliminate the overhead caused by the barriers, not only in the last 6 iterations of the loop, as proposed by Harris~\cite{harris2007optimizing}.

Our strategy is to use algebraic expressions to keep all the \textit{work-items} in the same execution step, maintaining its desired behaviour and algorithm correctness.

Consider the highly divergent code presented in Listing~\ref{code:ThreadDivergence2} (Section~\ref{subsec:AvoidingThreadDivergence}). Using a simple algebraic expression, it can be rewriten in order to completely eliminate the conditional statement and still return the right result of the comparison, as can be seen in Listing~\ref{code:Algebraic_if_then_else}.

\noindent
\begin{minipage}{\linewidth}
\begin{lstlisting}[frame=single,language=C,caption=Algebraic ``if-then-else''\label{code:Algebraic_if_then_else}]
 int smallestValue(int a, int b) {
  return (a < b) * a + (a >= b) * b;
 }
\end{lstlisting}
\end{minipage}

Note that the two boolean operations ($(a < b)$ and $(a >= b)$) are mutually exclusive, being interpreted internally by the compiler as 0 (false) or 1 (true). So, assuming that \textit{\textbf{a}} is smaller than \textit{\textbf{b}}, the result of the algebraic operation is \textit{\textbf{(1) * a + (0) * b}} which, ultimately, will return only the value of \textit{\textbf{a}}.

The same strategy can be applied to lines~\ref{ln049} to~\ref{ln054} of Listing~\ref{code:ParallelReductionBC1}, that represent the third step of the first stage. The new code is shown in Listing~\ref{code:AvoidingDivergences}, where $iLocalSize$ stores the number of active local work-items and $iLI$ represents the \textit{work-item's} local identifier.

\noindent
\begin{minipage}{\linewidth}
\begin{lstlisting}[frame=single,language=C,caption=Avoiding Divergences\label{code:AvoidingDivergences}]
 for (iPos = iLocalSize/2; iPos > 0; iPos >>= 1)
 {
  bFlag = iLI < iPos;
  scratch[iLI] += (bFlag)*(scratch[iLI + (bFlag)*iPos]);
 }
\end{lstlisting}
\end{minipage}

Here, in each iteration of the loop, $iPos$ is divided by 2 (\textit{iPos > > = 1}) and $bFlag$ is expanded to either 1 or 0, thus reducing by half the number of \textit{work-items} doing a useful job. If, for the current \textit{work-item}, the expression $iLI < iPos$ becomes true, then the expression in the last line will be interpreted as $scratch[iLI] += (1)*(scratch[iLI + (1)*iPos])$, ensuring that the value stored in position $iLI + iPos$ will be added to the value in position $iLI$. On the other hand, if the expression becomes false, it will be interpreted as $scratch[iLI] += (0)*(scratch[iLI + (0)*iPos])$, ensuring that the value in position $iLI$ will not be considered. Since all \textit{work-items} are always in the same step of computation -- doing exactly the same job (useful or not), independently of being in the same wavefront -- sync barriers are unnecessary.

\section{Computational Experiments}
\label{sec:computationalExperimentsPR}

\begin{figure}[ht]
  \centering
  \includegraphics[width=0.75\textwidth]{\sImagePath 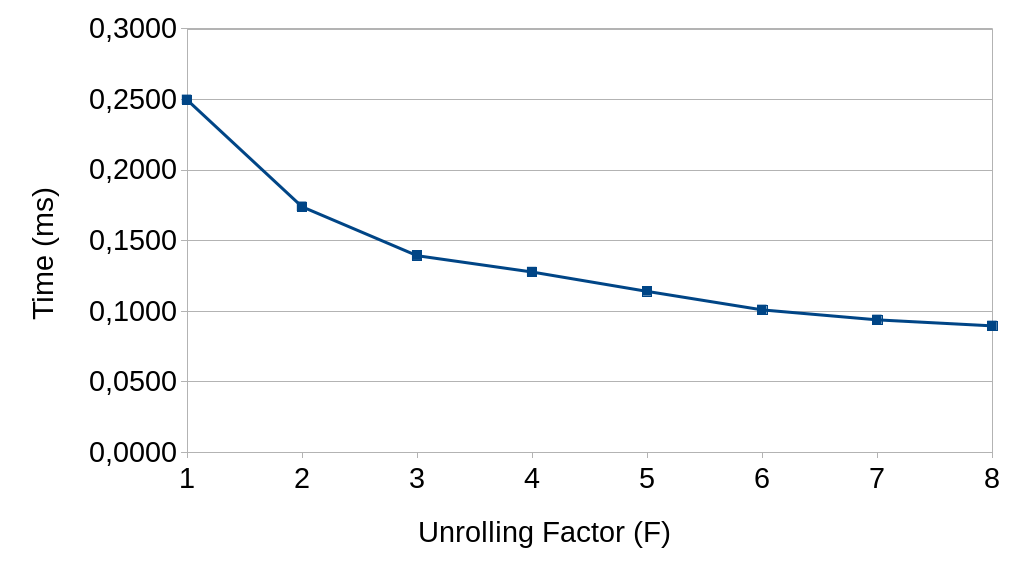}
  \caption{Chart of the parallel reduction execution times.}
  \label{fig:ChartParallelReductionExecutionTimes}
 \end{figure}

\begin{figure}[ht]
  \centering
  \includegraphics[width=0.75\textwidth]{\sImagePath 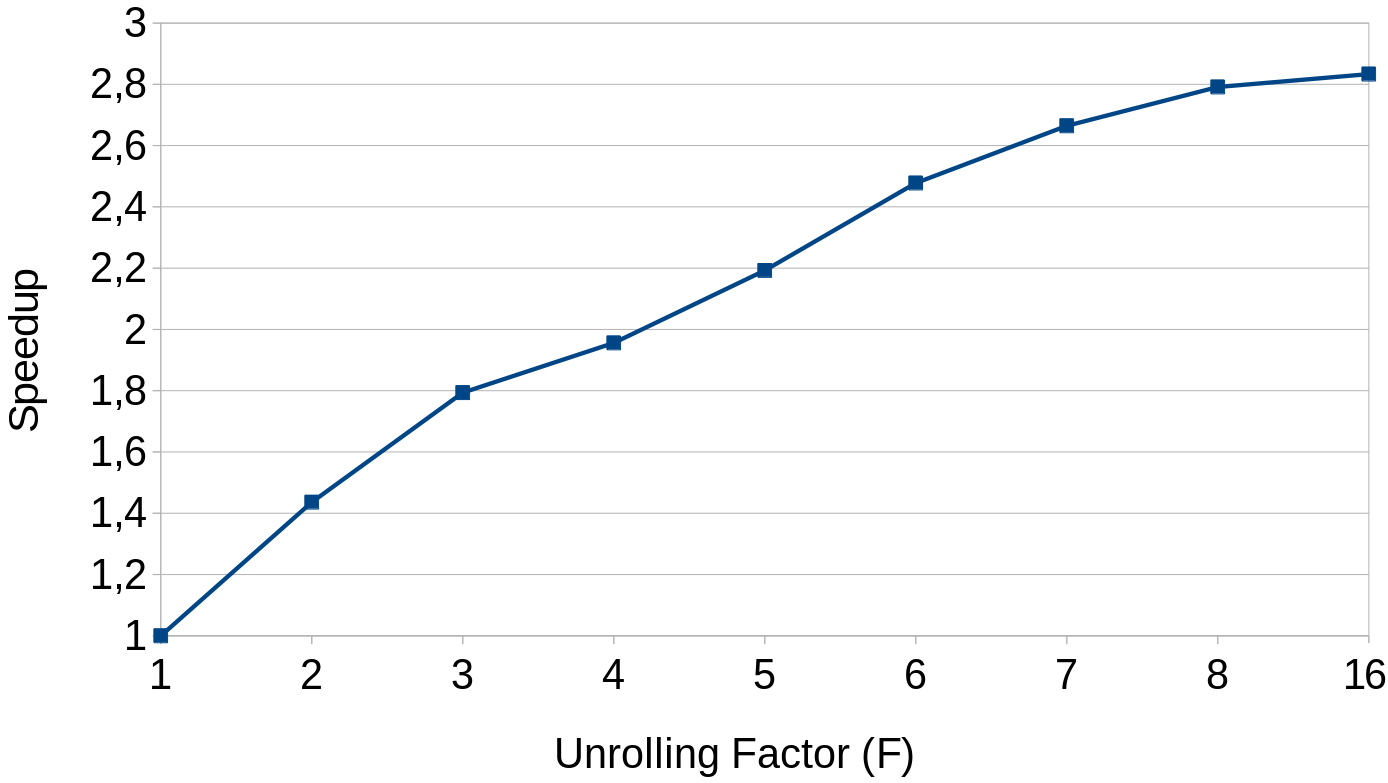}
  \caption{Chart of the parallel reduction speedup.}
  \label{fig:ChartParallelReductionSpeedup}
 \end{figure}

Table~\ref{table:ExecutionTimesTable} and Figures~\ref{fig:ChartParallelReductionExecutionTimes} and~\ref{fig:ChartParallelReductionSpeedup} represent the performance gains achieved against the algorithm described in~\cite{AMDParallelReduction2014}, where $F = 1$ is the runtime of the original code. The machine used in the tests was the same one presented in Section~\ref{sec:cc_results_gpu}.

All tests were run on two vectors, one of integers and one of single precision floating points, containing 5533214 elements. There were no measurable differences between the two vector types.

The times listed in Table~\ref{table:ExecutionTimesTable} were obtained with the OpenCL profiler CodeXL, version 2.0.12400.0, and are the averages of five consecutive executions for each \textbf{\textit{F}}.

As can be seen, these results show that the version of the algorithm with $F = 8$ reached a \textit{speedup} pretty close to 2.8x, when compared with the proposal of~\cite{AMDParallelReduction2014}. It may also be noted that such \textit{speedup} stabilizes around this value ($F = 16$ provided just over 1.5\% gain when compared to $F = 8$).

\begin{table*}[ht]
 \centering
  \begin{tabular}{|c|l|l|c|c|}
  \hline
    F   &  Time (ms)   &   Speedup        &   Memory Bandwidth (GB/s)  &   Bandwidth Usage (\%)  \\ \hline \hline
    1   &   0.249780   &   1              &   88.6094002722            &   26.63                 \\ \hline
    2   &   0.173930   &   1.4360949807   &   127.2515149773           &   38.24                 \\ \hline
    3   &   0.139260   &   1.7936234382   &   158.9318971708           &   47.76                 \\ \hline
    4   &   0.127700   &   1.955990603    &   173.3191542678           &   52.08                 \\ \hline
    5   &   0.113930   &   2.1923988414   &   194.2671464935           &   58.37                 \\ \hline
    6   &   0.100810   &   2.4777303839   &   219.5502033528           &   65.97                 \\ \hline
    7   &   0.093740   &   2.6646042245   &   236.1089822914           &   70.95                 \\ \hline
    8   &   0.089490   &   2.7911498491   &   247.3221142027           &   74.32                 \\ \hline
   16   &   0.088160   &   2.8332577132   &   251.0532667877           &   75.44                 \\ \hline
  \end{tabular}
 \caption{Parallel reduction execution times. New approach compared against Catanzaro's original code.}
 \label{table:ExecutionTimesTable}
\end{table*}

The same code was implemented in CUDA and tests were performed against the Kernel 7 of Harris presented in Section~\ref{subsec:MarkHarris}. The GPU used in the experiments was a Tesla C2075 with 6GB of memory. The architecture of such a video card provides 448 CUDA cores, a GPU clock of 575MHz and a shader clock of 1150Mhz. Its memory is clocked at 750MHz (3.0GHz effective).

The experiments employed the same two vectors containing 5533214 elements (integers and single precision floating points). Several values of the unrolling factor ($F$) were used in order to find the optimal value for such a video board. It was determined that up to $F=6$ the performance gains were substantial and, with $F \geq 8$, the gains were very discrete. According to this, all experiments were conducted using $F = 8$. Table~\ref{table:ExecutionTimesTableCUDAxOpenCL} presents the running time (in milliseconds) of both approaches and the percentage of performance (given by the formula $\frac{100 \ast T_{new}}{T_{k7}}$).

\begin{table*}[ht]
 \centering
  \begin{tabular}{|c|c|c|}
  \hline
    Time -- Kernel 7 & Time -- New Approach & \% of Performance  \\ \hline \hline
    0.17766 ms       & 0.17867 ms           & 99.4                \\ \hline
  \end{tabular}
 \caption[Parallel reduction execution times -- new approach compared against Harris' code.]{Parallel reduction execution times -- new approach (with unrolling factor equals to 8) compared against Harris' code.}
 \label{table:ExecutionTimesTableCUDAxOpenCL}
\end{table*}

% \subsection{Analysis of the Results}
% \label{subsec:AnalysisResultsPR}

\section{General Remarks}
\label{sec:GeneralRemarksParallelReduction}

Reduction operations are widely employed in many computational problems. This chapter showed how such operations can be performed in a parallel fashion using graphics processing units and detailed the main approaches for them nowadays.

% currently available.

All parallel reduction techniques currently in use suffer from some basic issues. Several only reach their peak performance by employing proprietary strategies and/or technologies, what ends up limiting their use to the platform for which they were designed. Others, though generic, do not adopt certain procedures that could increase their performance without loss of generality.

The strategy presented here combines the best of both worlds: It is generic enough to be used with both CUDA and OpenCL and can run on hardware of the two major GPU manufacturers with minimal changes, just being adapted to the particularities of each platform. The implemented code, besides simpler, offered a performance equivalent to the best strategy described by Harris~\cite{harris2007optimizing}.

A good performance of this routine is essential for the efficient execution of the macroscopic urban traffic assignment algorithm described in Chapter~\ref{chapter:traffic_assignment_problem}, since it is used on two occasions: in the computation of shortest paths and in the golden ratio method.

\bibliographystyle{abbrv}
% \bibliography{/home/walid/DiscoD/Diversos/DOCS/Doutorado/Estudo_Dirigido/references}
\bibliography{references}
\label{ref-bib}

\end{document}